\documentclass[12pt]{iopart}

\usepackage{iopams}
\usepackage{graphicx}
\usepackage{color}
\begin{document}

\title[Double-point defects in SC]{Propagating and evanescent properties of double-point defects in Sonic Crystals}

\author{V. Romero-Garc\'ia$^{1,2}$, J.V. S\'anchez-P\'erez$^1$ and L.M. Garcia-Raffi$^3$.}

\address{$^1$ Centro de tecnolog\'ias f\'isicas: Ac\'ustica, Materiales y Astrof\'isica, Universidad Polit\'ecnica de Valencia, Camino de Vera s/n, 46022 Valencia}
\address{$^2$ Instituto de Ciencia de los Materiales. Consejo Superior de Investigaciones Cient\'ificas. Sor Juana Inés de la Cruz, 3, Cantoblanco, 28049, Madrid, Spain.}
\address{$^3$  Instituto Universitario de Matem\'atica Pura y Aplicada, Universidad Polit\'ecnica de Valencia, Camino de Vera s/n, 46022 Valencia}
\ead{virogar1@mat.upv.es}

\begin{abstract}
Complex Band Structures and Multiple Scattering Theory have been
used in this paper to analyze the overlapping of the evanescent
waves localized in point defects in Sonic Crystals. The Extended
Plane Wave Expansion (EPWE) with supercell approximation gives the
imaginary part of the Bloch vectors that produces the decay of the
localized modes inside the periodic system. Double-cavities can
present a coupling between the evanescent modes localized in the
defect, showing a symmetric or antisymmetric modes. When point
defects are close, the complex band structures reveal a splitting
of the frequencies of the localized modes. Both the real part and
the imaginary values of $k$ of the localized modes in the cavities
present different values for each localized mode, which gives
different properties for each mode. Novel measurements, in very
good agreement with analytical data, show the experimental
evidences of the symmetric and antisymmetric localized modes for a
double-point defect in Sonic Crystals (SC). The investigation on
the localization phenomena and the coupling between defects in
periodic systems has a fundamental importance both in pure and in
applied physics.
\end{abstract}


\maketitle

\section{Introduction}
Periodic distributions of elastic scatterers in an elastic host
medium with different physical properties are known as Phononic
Crystals (PC) \cite{Sigalas93, Kushwaha93} and they are the elastic
analogous of the well-known Photonic Crystals \cite{Yablonovitch,
John87}. If one of the materials in PC is a fluid, then the system
is called Sonic Crystal (SC) \cite{Martinez05}. All of these
systems present interesting physical properties and recently they
have received increasing attention mainly due to the great
number of applications in several branches of physics and
engineering \cite{Khelif04, Joannopoulus08, Sigalas05}. One of the
most important properties of these inhomogeneous materials are the
so-called Band Gaps (BG): frequency ranges where waves do not
propagate through the periodic system. The existence of these BG
leads to a several applications, for instance, in the case of SC
as acoustic filters  \cite{Sigalas97, Sigalas98}, acoustic
barriers \cite{Sanchez02} or waveguides \cite{Vasseur08}, among
others.

In periodic systems Bloch's theorem and Fourier expansion of the
periodic physical properties, transform the acoustic wave equation
in an eigenvalue problem. The eigenfrequencies $\omega(k)$ for
each Bloch's vector $k$ inside the irreducible part of the first
Brillouin Zone constitute the band structure. This methodology is
usually called Plane Wave Expansion (PWE) \cite{Kushwaha94} and it
can be used to obtain the so-called band structures, i.e. the
propagating modes through the periodic system. The band structures
reveal that BG are ranges of frequencies where no real $k$ exists. It has been shown that the eigenvalues of the problem have
real value for the case of SC \cite{Halevi95}.

One of the most important properties of the periodic structures is
the emergence of the localized modes within the BG when a point
defect is introduced \cite{Sigalas97, Li05}. A widely technique
used in the literature to obtain the effect of the creation of
point defects in crystals is the supercell approximation in
PWE \cite{Sigalas98, Wu01, Zhao09}. This approximation gives
information only about the propagation nature of the localized
modes in the point defects. In these cases when periodicity is
broken or when SC have finite size, evanescent modes inside the
periodic system may appear. Localized modes or modes inside the BG
are characterized by its evanescent behaviour \cite{Joannopoulus08,
Engelen09, Romero10}. Then a more accurate analysis is needed to
characterize all the properties of the modes inside the periodic
system.

A wave impinging on a complete periodic system with a given
frequency $\omega$ inside the BG, is characterized by complex
valued wave numbers $k(\omega)$ which represent the
multiexponential decay of the evanescent mode inside the periodic
system \cite{Engelen09}. Recent works  \cite{Hsue05, Laude09,
Romero10_2} show an extension of the Plane Wave Expansion
(Extended Plane Wave Expansion (EPWE)) obtaining the complex part
of the Bloch's waves revealing that the decay of the modes inside
the BG grows as the frequency reaches the center of the BG. 
In this sense a localization
factor has been defined recently to show this
behaviour \cite{Wang09}. The localization factor can be also
related with recent results that show that, although the decay of
these localized modes is multi-exponential, it can be approximated
by an exponential like decay considering only the first harmonic
of the Bloch waves in SC made of rigid cylinders \cite{Romero10}.

  On the other hand,  Sainidou et al. \cite{Sainidou05} have introduced a novel extension of the Multiple Scattering Theory \cite{Psarobas00} for analyzing slabs which consist of slices of different material as long as the periodicity parallel to the surface of the slab is preserved. The method, called Layer Multiple Scattering (LMS) allows to study the scattering problem of slabs that are finite in the direction parallel to the surface of the slab, but infinite in the normal directions to this surface. Alternatively, one can use this method to calculate the complex phononic band structures of an infinite
crystal, associated with a given crystallographic plane. In this case the method provides the propagating and
evanescent Bloch waves of the elastic field in the given crystal, corresponding to a given $k$ and a given frequency. LMS has been used to analyze the guidance and quasiguidance of elastic waves in a glass plate coated on one side with a periodic monolayer of
polymer spheres, immersed in water, observing the dispersion diagrams of the interacting modes of the composite slab \cite{Sainidou06}.

The goal of the paper is to analyze the three main characteristics
of defect modes in SC: splitting, symmetry vibrational patterns
and evanescent decay of the modes. In addition to the PWE, to do
this study we have used the EPWE with supercell approximation
because it is fundamental for the complete understanding of the
localized modes. We present the explicit matrix formulation of the
supercell approximation in the EPWE for $N_p$ point defects. From
the complex and real band structures we observe the splitting, and
the evanescent behaviour of the localized modes inside the BG
around the defect. We analyze the localized modes inside
multi-point defects, specially in the double point defect case.

MST in finite SC is used to analyze the
vibrational patterns of the localized modes in a double point
defect. In this case, when the distance between both
defects is low enough, it appears a symmetric and antisymmetric
vibrational modes similar to the case of a system formed by two
masses and three springs, or to the Zeeman effect in the atomic
spectra \cite{Li05}. Novel experimental data that are in good
agreement with theory show for first time the symmetry of the
vibrational patterns of the localized modes in such a double point
defect. Moreover we observe the decay of the localized modes
outside the double-point defect in good agreement with the results
obtained by EPWE with supercell approximation.

The paper is organized as follows. First of all we show the main
ingredients of the Extended Plane Wave Expansion as well as the
explicit matrix formulation of the problem and the extension for a
supercell with $N_p$ point defects. After that, it is shown the
numerical, analytical and experimental results of a double-point
defect, showing the complete explanation of the splitting, the
symmetry of the vibrational patterns and the decay of the
localized modes. Finally we show a summary as well as the main
conclusions of the work.

\section{Extended PWE with supercell approximation}
\label{sec:EPWE} The analysis of the propagating modes can be done
by the $\omega(\vec{k})$ formulation, where the existence of BG is
indicated by the absence of bands in determined ranges of
frequencies. The mechanism of creation of BG in finite crystals
could be understood by the evanescent behaviour of the modes
inside it. At a given frequency $\omega$ inside the BG, the
evanescent wave is characterized by a complex valued Bloch vectors
$\vec{k}(\omega)$ that characterize the decay of the mode inside
the periodic structure. Based on the work of Hsue et al.
 \cite{Hsue05} a recent work by Laude et al. \cite{Laude09} shows
the calculation of complex band structure for phononic crystals.
Recently, this work has been extended for the case of SC for
calculations using the supercell approximation \cite{Romero10_2},
which is specially indicated for SC with point defects. In this
section we present the explicit matrix formulation of the Extended
Plane Wave Expansion with supercell approximation to calculate the
properties of SC with $N_p$ point defects inside a supercell. We
must take into account that PWE needs a low interaction between
supercells.

$\omega(k)$ methods are characterized by the next eigenvalue
problem:
\begin{eqnarray}
\sum_{\vec{G'}}((\vec{k}+\vec{G})\sigma_k(\vec{G}-\vec{G'})(\vec{k}+\vec{G'})-\omega^2\eta(\vec{G}-\vec{G'}))p_{\vec{k}}(\vec{G'})=0.
\label{eq:eigenproblem}
\end{eqnarray}
Where $\vec{G}$ is the 2D reciprocal-lattice vector, $k$ is the
Bloch vector and $p_k$ is the pressure. Equation
(\ref{eq:eigenproblem}) constitutes a set of linear, homogeneous
equations for the eigenvectors $p_{\vec{k}(\vec{G})}$ and the
eigenfrequencies $\omega({\vec{k}})$. We obtain the band
structures letting $\vec{k}$ scan the irreducible part of the
first Brillouin zone.

Equation (\ref{eq:eigenproblem}) can be expressed by the next
matrix formulation  \cite{Kushwaha94}
\begin{eqnarray}
\label{eq:matricial} \sum_{i=1}^3\Gamma_i\Sigma\Gamma_i P=\omega^2
\Omega P,
\end{eqnarray}
where i=1,2,3, and
\begin{eqnarray}
\Sigma=\left( \begin{array}{ccc}
\sigma(\vec{G}_1-\vec{G}_1) & \ldots & \sigma(\vec{G}_1-\vec{G}_{N\times N}) \\
\vdots & \ddots & \vdots \\
\sigma(\vec{G}_{N\times N}-\vec{G}_1) & \ldots & \sigma(\vec{G}_{N\times N}-\vec{G}_{N\times N})\\
\end{array}
\right),\label{eq:Sigma_matrix}\\[0.1 cm]
\Omega=\left( \begin{array}{ccc}
\eta(\vec{G}_1-\vec{G}_1) & \ldots & \eta(\vec{G}_1-\vec{G}_{N\times N}) \\
\vdots & \ddots & \vdots \\
\eta(\vec{G}_{N\times N}-\vec{G}_1) & \ldots & \eta(\vec{G}_{N\times N}-\vec{G}_{N\times N})\\
\end{array}
\right),\label{eq:eta_matrix}\\[0.1 cm]
P=\left(\begin{array}{c}
P(\vec{G}_1)\\
\vdots\\
P(\vec{G}_{N\times N})\\
\end{array}
\right),
\end{eqnarray}
where $\vec{G}=(G_1, G_2, G_3)=(2\pi m/a_1,2\pi n/a_2, 0)$ for the
case of the 2D square arrays . If we chose $m=n=(-M,\ldots,M)$,
the size of the previous matrices is $N\times N=(2M+1)\times
(2M+1)$.

From equation (\ref{eq:matricial}) we define the next vector,
\begin{eqnarray}
\Phi_i=\Sigma\Gamma_iP.
\end{eqnarray}
With this definition we can reformulate the eigenvalue problem
(\ref{eq:matricial}) as the equations system
\begin{eqnarray}
\Phi_i=\Sigma\Gamma_iP\nonumber\\
\omega^2\Omega P=\sum_{i=1}^3\Gamma_i\Phi_i.
\end{eqnarray}
In order to obtain an eigenvalue problem for $\vec{k}(\omega)$, we
write $\vec{k}=k\vec{\alpha}$, where $\vec{\alpha}$ is a unit
vector. Then $\Gamma_i$ matrix can be written as
\begin{eqnarray}
\Gamma_i=\Gamma_i^0+k\alpha_iI,
\end{eqnarray}
where $I$ is the identity matrix, and
\begin{eqnarray}
\Gamma_i^0=\left(
\begin{array}{cccc}
G_i & 0 & \ldots & 0 \\
0 & G_i & \ldots & 0 \\
\vdots & \vdots & \ddots & \vdots\\
0 & \ldots & \ldots & G_i  \end{array}
\right),\label{eq:Gamma_matrix_b} \\[0.5cm]
 \alpha_i=\left(
\begin{array}{cccc}
\alpha_i & 0 & \ldots & 0 \\
0 & \alpha_i & \ldots & 0 \\
\vdots & \vdots & \ddots & \vdots\\
0 & \ldots & \ldots & \alpha_i  \end{array}
\right).\label{eq:alpha_matrix_b}
\end{eqnarray}

Then, equation (\ref{eq:matricial}) can be written in the form of
(\ref{eq:matricial_complex}), where
$\Phi'=\sum_{i=1}^3\alpha_i\Phi_i$.
\begin{eqnarray}
\left(
\begin{array}{cc}
\omega^2\Omega -\sum_{i=1}^3\Gamma_i^0\Sigma\Gamma_i^0 & 0  \\
-\sum_{i=1}^3\Sigma \Gamma_i^0 & I\end{array} \right) \left(
\begin{array}{c}
P \\
\Phi' \end{array}\right)=k \left(
\begin{array}{cc}
\sum_{i=1}^3\Gamma_i^0\Sigma\alpha_i & I  \\
\sum_{i=1}^3\Sigma \alpha_i & 0\end{array} \right) \left(
\begin{array}{c}
P\\
\Phi'\end{array} \right) \label{eq:matricial_complex}
\end{eqnarray}

Equation (\ref{eq:matricial_complex}) represents a generalized
eigenvalue problem with $2N$ eigenvalues $k$, with possibly
complex values. Complex band structures on the incidence direction
$\vec{\alpha}$ have been obtained by solving the eigenvalue
equation for a discrete number of frequencies and then sorted by
continuity of $k$. In contrast to the $\omega(\vec{k})$ method, in
this formulation the periodicity is not relevant and $k(\omega)$
does not follow the first Brillouin zone.

We consider a SC with primitive lattice vectors $\vec{a}_i$ ($i=1,2,
3$). The supercell is a cluster of $n_1a\times n_2a\times n_3a$
scatterers periodically placed in the space. Then, the primitive
lattice vectors in the supercell approximation are
$\vec{a'}_i=n_i\vec{a}_i$, and the complete set of lattices in the
supercell approximation is $\{R'|R'=l_i\vec{a'}_i\}$, where $n_i$
and $l_i$ are integers. The primitive reciprocal vectors are then
\begin{eqnarray}
\vec{b'}_i=2\pi \frac{\varepsilon_{ijk}\vec{a'}_j\times
\vec{a'}_k}{\vec{a'}_1\cdot(\vec{a'}_2\times \vec{a'}_3)}
\end{eqnarray}
where $\varepsilon_{ijk}$ is the three-dimensional Levi-Civita
completely anti-symmetric symbol. The complete set of reciprocal
lattice vectors in the supercell is
$\{\vec{G}|\vec{G}_i=N_i\vec{b'}_i\}$ where $N_i$ are integers.

The density $\rho_i$ and the bulk modulus $B_i$, are the physical
properties involved in the wave equation and using the Fourier
expansion and the geometry of the system, they can be expressed in
terms of the structure factor for the PWE (EPWE) as well as for
the PWE (EPWE) with supercell approximation. The index $i=(h,c)$
represents the host medium and the scatter respectively. The
filling fraction of a cylinder in a supercell is $f=\pi r^2/A$,
where $A$ is the area occupied by the supercell. If we consider
that $\beta_i$ represents the values $(\rho_i^{-1}, B_i^{-1})$,
and that the supercell has $N_c$ cylinders organized in an array
of size $n_1a\times n_2a$ then
\begin{eqnarray}
  \beta(\overrightarrow{G})= \left\{ \begin{array}{ll}
        \beta_{c}N_cf+\beta_{h}(1-N_cf)& \mbox{if $\overrightarrow{G} = \overrightarrow{0}$}\\
        \left(\beta_{c}-\beta_{h}\right)F(\overrightarrow{G}) & \mbox{if $\overrightarrow{G} \neq \overrightarrow{0}$}
        \end{array}\right.
\end{eqnarray}
where $F(\overrightarrow{G})$ is the structure factor of the
supercell.

In this approximation the structure factor of the supercell has to
be computed taking into account the size of the supercell. If we
consider a 2D SC with cylindrical scatterers with radius $r$ and
size of the supercell $n_1a\times n_2a$, the structure factor of
the supercell is expressed by
\begin{eqnarray}
F(\vec{G})=\sum_{i=-(n_1-1)/2}^{(n_1-1)/2}\sum_{j=-(n_2-1)/2}^{(n_2-1)/2}e^{\imath(ia|\vec{G}_1|+ja|\vec{G}_2|)}P(\vec{G})
\end{eqnarray}
where
\begin{eqnarray}
P(\vec{G})=\frac{2f}{Gr}J_{1}(G).
\end{eqnarray}
where $a$ is the lattice constant inside the supercell and
$G=|\vec{G}|$.

Previous equations show the expressions for the approximation of
complete supercell. If the supercell presents $N_p$ point defects
at the sites labelled by  $(l_s,m_s)$ in the periodic system, with
$s=1,...,N_p$, then the Fourier coefficients of the expansions of
the physical parameters involved in the problem follow the next
equation
\begin{eqnarray}
  \beta(\overrightarrow{G})= \left\{ \begin{array}{ll}
        \beta_{c}(N_c-N_p)f+\beta_{h}(1-(N_c-N_p)f)& \mbox{if $\overrightarrow{G} = \overrightarrow{0}$}\\
        \left(\beta_{c}-\beta_{h}\right)F(\overrightarrow{G}) & \mbox{if $\overrightarrow{G} \neq \overrightarrow{0}$}
        \end{array}\right.
\end{eqnarray}
The structure factor of such a supercell with $N_p$ point defects
is

\begin{eqnarray}
\label{eq:multi_point}
F(\vec{G})=\left(\sum_{i=-(n_1-1)/2}^{(n_1-1)/2}\sum_{j=-(n_2-1)/2}^{(n_2-1)/2}e^{\imath(ia|\vec{G}_1|+ja|\vec{G}_2|)}-\sum_{s=1}^{N_p}e^{\imath(l_{s}a|\vec{G}_1|+m_{s}a|\vec{G}_2|)}\right)P(\vec{G}).
\end{eqnarray}

The interaction of the defect points in the supercell
approximation must be as low as possible between the neighboring
supercells in order to decrease the overlap between defects, thus
the size of the supercell should be big enough to place the point
defects separated in consecutive supercells.

Introducing the previous expressions in the matrices of both the
PWE (\ref{eq:matricial}) or the EPWE (\ref{eq:matricial_complex})
we can calculate the real and complex band structures. In the
present paper we analyze the case of a double-point defect in a
square array at sites $(1,0)$ and $(-1,0)$ in a supercell of
$11a\times 11a$. In this situation the distance between defects is
equal to $2a$, and between two double-point defects in different
supercells is equal to $20a$.

\section{EPWE Results: Localized modes}
\label{sec:results}

Since Sigalas et al. \cite{Sigalas97} studied the defect mode
produced by a point defect in periodic structures, several kinds
of defects have been analyzed in the last years, showing in all
cases the localization of sound for frequencies inside the
BG \cite{Li05, Wu03, Zhong05}. Experimental and numerical analysis
of the localization in a point defect considered as a cavity
inside the SC have been reported recently by Wu et al.
 \cite{Wu09PB, Wu09PLA} and by Zhao et al.  \cite{Zhao09, Zhao09_2}
showing the dependence of the localization on the size of the
crystal and on the filling fraction (the bigger the size and the
filling fraction the bigger the localization in the cavity).
Moreover, when we consider two point defects, coupling between
localized modes in each defect-point is possible \cite{Khelif03, Russell03}. An accurate interferometric set up has been used by Russell et al. \cite{Russell03} for observing the coupled states
in a double-point defect, noting the evidences for odd and even symmetry trapped states in new class of ultra-efficient photosonic devices in which both sound and light
are controlled with great precision and their interactions enhanced.
For the case of double-point defect, the bigger the distance between cavities the lower
coupling between defect points.

In this section we show novel results about the imaginary part of
the Bloch vector for the localized modes inside the SC with
multi-point defects. The localization of waves inside these
defects is mainly characterized by three properties. First, the
modes are separated in the frequency domain, i.e., there is a
splitting of the localization frequency if the point defects are
close enough. Second, the modes present symmetries in the
vibrational pattern depending on the number of vacancies in the
crystal. Third, the localized modes are evanescent and they decays
outside the defect but inside the SC. Without loss of generality
we show results of a double point defect in very good agreement
with the experimental data, showing the symmetric and
antisymmetric vibrational pattern of the localized modes. Evidently, the oscillation modes of $N$-point defects with $N>2$ will present more complicated vibrational patterns than the ones appearing in the double-point defect, then they cannot be classified into such simple modes as symmetric and antisymmetric ones.

The Complex and Real band structures reveal that the values of $k$ for the localized modes are characterized by a real value of $k$ and it is related with the localization frequency. However, this localized mode presents evanescent behaviour out of the defect because it is surrounded by a perfect periodicity, thus the excited mode in the surrounding crystal by the localized mode presents an imaginary $k$ which is related with the evanescent behaviour of the mode outside the point defects \cite{Romero10}. The rest of modes inside the BG only
present imaginary part, then they are killed inside the crystal
because of their evanescent behaviour.

\subsection{Splitting of localized modes}
\label{sec:splitting}

In order to analyze the splitting of the localized modes we have
calculated the Real and the Complex Band Structures of a SC with a
double point defect with the EPWE with the supercell approximation
using a supercell of size $11a\times 11a$. We consider a 2D SC
consisting of PVC cylinders of radius $r$ in air background
arranged in square lattice with lattice constant $a$. The material
parameters employed in the calculations are
$\rho_{air}=1.23$kg/$m^3$, $\rho_{PVC}=1400$kg/$m^3$,
$c_{air}=340$m/s and $c_{PVC}=2380$m/s. We consider a filling
fraction $f=\pi r^2/a^2\simeq0.65$. For the calculations we have
used $N=(2\cdot15+1)^2=961$ plane waves. Several calculations have been carried out in order to obtain a good convergence of the solution. This number of plane waves is bigger than the used in previous works \cite{Laude09} and it provides a good convergence of the solution of the eigenvalue problem.

\begin{figure}
\begin{center}
\includegraphics[width=90mm,height=85mm,angle=0]{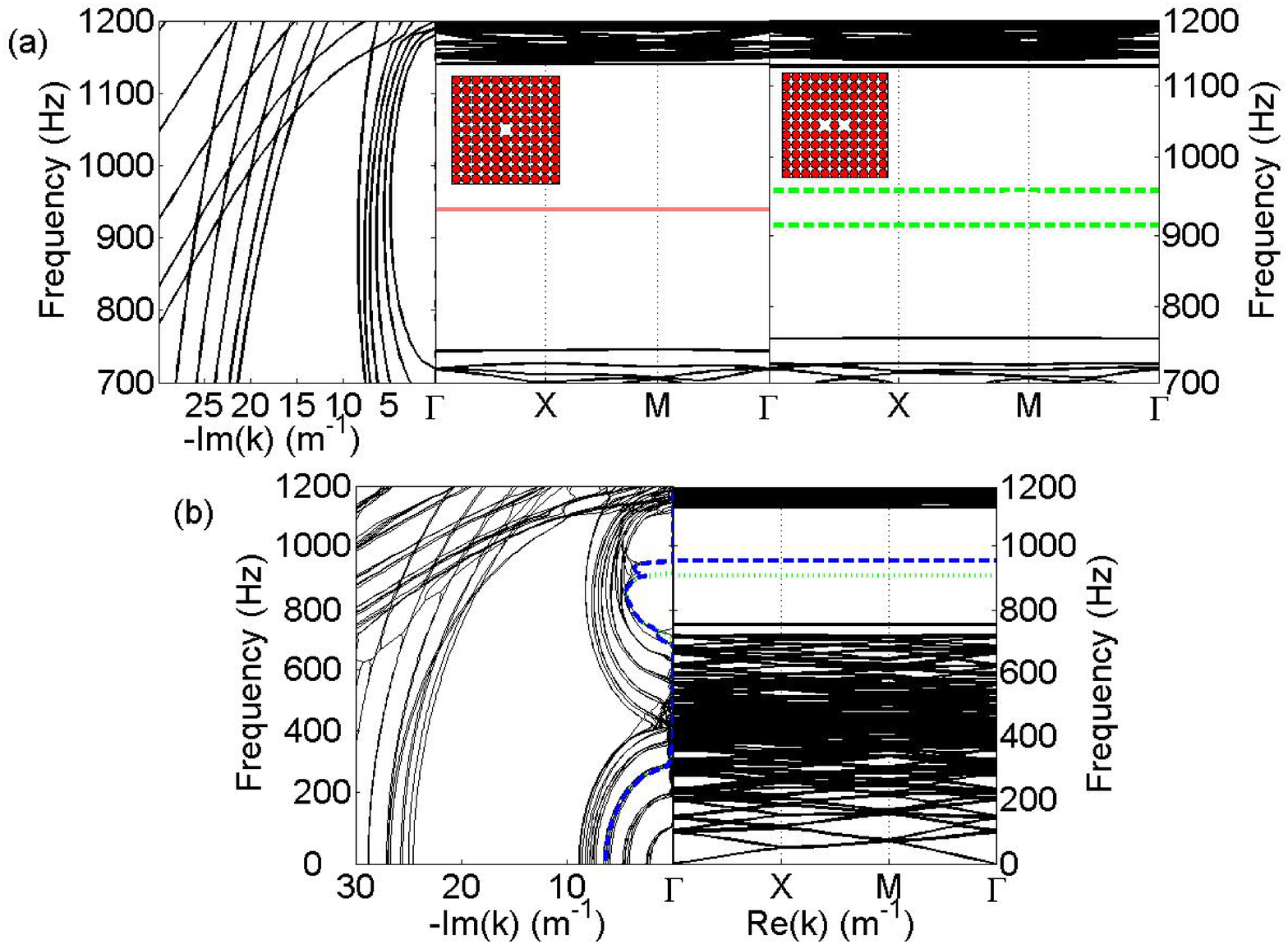}%
\caption{\label{fig:EPWE_figure} (Color Online) Real and Complex
band structures for a SC with point defects. (a): Complex band
structure of a complete SC calculated by EPWE with supercell
approximation (left). Band structures calculated by PWE with
supercell approximation of a SC with a point defect, the
continuous red line represents the defect mode (center). Band
Structures for a SC with a double-point defect, dashed green line
represents the defect modes of a double point defect (right).
Insets show the supercell used in the calculations. (b): Complex
and real band structure of a double point-defect.}
\end{center}
\end{figure}

A mode within the BG in an infinite SC without defects is characterized by a pure
imaginary value of $k=\imath k_{im}$ (where
$k_{im}=Im(k)$) \cite{Laude09, Romero10, Wang09}. In Figure
\ref{fig:EPWE_figure}a (left panel) we can observe the dependence
of the $Im(k)$ on $k$ for a complete SC within the BG for the
$\Gamma$X direction. We can observe a maximum value of $Im(k)$ for
the frequency in the midgap (926 Hz), that means the imaginary
part of the wave number for frequencies inside the BG grows with
values of frequency closer to the center of the BG and disappears
at the edges of the BG, i. e., the rate of decay is bigger for
frequencies closer to the center of the BG \cite{Joannopoulus08,
Wang09, Romero10}. Modes within the BG decay inside the SC because
of their evanescent behaviour \cite{Romero10}.

In contrast to the modes in BG, localized modes can travel up to
the point defect where the wave is localized. Figure
\ref{fig:EPWE_figure}a (central and right panels) represents the
real band structures calculated by PWE with supercell
approximation for both a SC with a point defect (central) and a SC
with a double-point defect (right). We can observe the localized
mode generated by a point defect in a SC at the frequency
$\nu_0=932$ Hz whereas the frequencies of the localized modes of a
double point defect have been split (right panel of the Figure
\ref{fig:EPWE_figure}a). The frequencies of the two localized
modes due to the double-point defect split around the localized
mode of a single defect: One with lower frequency, $\nu_1=910$ Hz,
than the corresponding frequency of the localized mode in a single
defect, and another one, $\nu_2=958$ Hz, with higher frequency
than the single defect. This phenomenon is analogous to the
splitting of the degenerate atomic levels in diatomic molecules.

The splitting in two peaks may be understood qualitatively by
considering that the double cavity in the double-point defect are
coupled forming a large cavity with two resonant frequencies. When
a wave with one of these frequencies impinges on the crystal from
outside, the double point defect is sawn and the wave penetrates
into the cavity. Then, the borders of the cavity act as a perfect
mirrors producing the localization of the wave inside the cavity.
This results in a coupling inside the double-point defect
producing two localized modes depending on the distance between
the point defects \cite{Li05, Khelif03, Zhao09}.

Because the splitting depends on the distance and on the shape
of the multi-point defect, one can study the vibrational patterns
that appear inside the multi-point defect, analyzing the
differences in frequency of the localized modes. The factor
$(\nu_{1}-\nu_2)/\nu_0$ indicates how the splitting will be
produced. For big values of this factor, one can expect modes
separated in frequencies (as many as single-point defect
constituting the multi-point defect), whereas a small factor
represents a weak overlap between the point defects in the
multi-point defect, which produces narrow splitting.

The complex band structures give us additional information about
the properties of the localized modes. Figure
\ref{fig:EPWE_figure}b represents the complex (left panel) and the
real (right panel) band structures for a SC with a double-point
defect.
For each localized mode a determined imaginary $k$ becomes in a
pure real value, in good agreement with the results of the PWE
with supercell approximation. This real part is related with the
wave vector for the localization frequency
whereas the imaginary part of the localized mode 
is related with the rate of decay outside the defect but inside
the SC.
 As we have seen, the localized modes in the double point defect
are distributed around the localized mode of a single point
defect. However, the localized mode of the single defect point
appears a little above the midgap frequency (926 Hz). Thus it is
expected that the imaginary part of the localized modes of
double-point defects present
different values for each mode and, as a consequence, 
each mode presents different evanescent behaviour outside the
defect (see Figure \ref{fig:EPWE_figure}b). This prediction of the
EPWE will be used for distinguishing experimentally the symmetric
respect to the antisymmetric modes.

\subsection{Symmetric and antisymmetric modes} \label{sec:num_exp}
The previous discussion about the splitting of the modes in
multi-point defects does not provide information about how the
modes are localized or what is the distributed field inside the
double cavity. This will be discussed now.

The results obtained by the PWE or EPWE for the localized modes could be used to plot the modal shapes for the defect modes using the eigenvectors. However, these modal shapes do not take into account the effect of the finite size of the crystal. Thus, for comparing with experimental data corresponding to a SC of finite size, in this Section, we have calculated the modal shapes inside the double defect using Multiple Scattering Theroy (MST). MST provides complementary information respect to the one provided by EPWE in the case of the infinite structures.

MST \cite{Linton01, Chen01} has been
used to analyze the pressure field inside a SC with point defects. A SC of $7a\times 5a$ size with $a=0.22$m of rigid cylinders with
radius $r=0.1$m is considered in this Section. We have generated a
double-point defect with individual defect points separated a
distance $d=2a$. We have considered this size because of the
experimental restrictions, and to be able to compare both
theoretical and experimental data.

For the crystal considered in this Section the frequencies of the
localization modes differs a bit respect to the calculated by PWE
and EPWE with supercell approximation. We have to take into
account that in this case we consider a finite crystal, and as it
has been shown in the literature, the localization frequencies
depend on the size of the crystal as well as the filling fraction,
and the amount of rows around the defect. In this case, the
localization frequency for the antisymmetric mode is $\nu_1=940$
Hz, and for the symmetric mode is $\nu_2=895$ Hz. These
frequencies represent the maxima values of the acoustic spectra
inside the point defects. Note the small difference respect to the
obtained by PWE and EPWE.

\begin{figure}
\begin{center}
\includegraphics[width=90mm,height=45mm,angle=0]{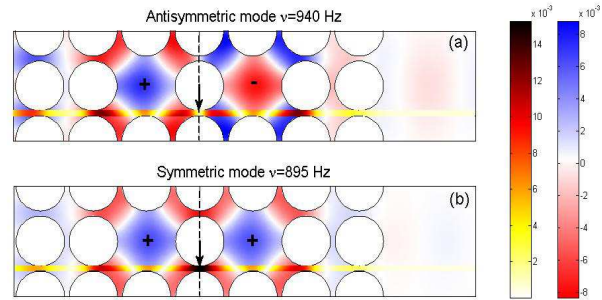}%
\caption{\label{fig:maps_degenerated} (Color Online) Pressure maps
of a double-point defect separated a distance $d=2a$. The value
$|p|$ between two rows of the SC containing the point defects is
also plotted. Pressure map of the antisymmetric (a) and symmetric
(b) coupling of the localized modes inside the double-point
defect. Arrow represents the values of $|p|$ in the midpoint (dashed line) between two
rows of cylinders containing the double point defect.}
\end{center}
\end{figure}

The pressure fields calculated by MST inside the SC with a
double-point defects for the localization frequencies are shown in
the Figure \ref{fig:maps_degenerated}. We can observe that the
pressure field for the mode with high frequency has an
antisymmetric pattern Figure \ref{fig:maps_degenerated}a, whereas
the pressure field for the mode with low frequency has a symmetric
pattern Figure \ref{fig:maps_degenerated}b.

In Figure \ref{fig:maps_degenerated} one can also observe the
values of $|p|$ for the space between two rows of the SC
containing the double-point defect. The vibrational patterns of
the defect modes in double-point defect are characterized respect
to a symmetry plane (see dashed line in Figure
\ref{fig:maps_degenerated}) situated just in the midpoint between
the two defects in the double-point cavity. There is a symmetric
mode and an antisymmetric mode respect this plane. The symmetric
vibrational mode is characterized by a vibration in phase of the
pressure field in each point defect, whereas the antisymmetric
mode is characterized by a vibration of the pressure field with
opposition of phase. Because of these properties the point just in
the symmetry plane presents different values of $|p|$ for each
localized mode (see arrows in Figure \ref{fig:maps_degenerated}).
For the antisymmetric mode (Figure \ref{fig:maps_degenerated}a)
one can observe a minimum value of $|p|$ at this point, whereas
one can find a maximum value for the symmetric mode (Figure
\ref{fig:maps_degenerated}b).

Experimental results inside the SC have been performed in an
echo-free chamber of dimensions $8\times 6\times 3$m$^3$. In order
to obtain the experimental dependence of the pressure all along
the SC, we have measured the pressure field in several points
between two rows of the SC containing the double-point defect. The experimental SC has been made of 1m long cylinders of PVC.
The size of the SC considered in this work has the adequate
dimensions to be capable to introduce the microphone between rows.
The microphone used is a prepolarized free-field 1/2" Type $4189$
B\&K. The diameter of the microphone is $1.32$cm, which is
approximately $0.06a$, so the influence over the pressure field
measured is negligible.

With our system 3DReAMS (3D Robotized e-Acoustic Measurement
System) it is possible to sweep a 3D grid of measuring points
located at any trajectory inside the echo-free chamber. Motion of
the robot is controlled by NI-PCI 7334 . We have analyzed the
absolute value of the sound pressure between two rows of the SC
moving the microphone in steps of $1$cm.

\begin{figure}
\begin{center}
\includegraphics[width=90mm,height=45mm,angle=0]{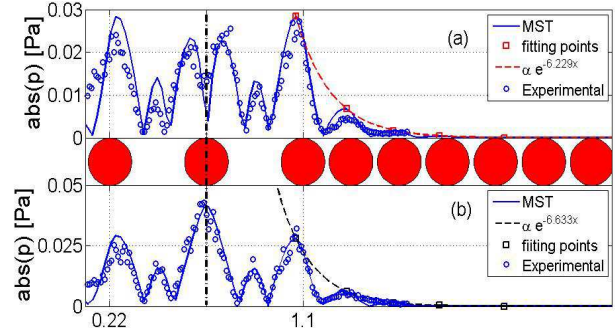}%
\caption{\label{fig:profiles_degererated} (Color Online) Numerical
(continuous line) and experimental (open circles) profile of the
$|p|$ between two rows containing the double-point defect (see
Figure \ref{fig:maps_degenerated}). (a) Antisymmetric mode
($\nu=940$Hz) and (b) symmetric mode ($\nu=895$Hz). Dashed line
represents the exponential decay of the localized modes outside
the double-point defect fitted from the maxima values of the
analytical data represented with square open points. The border of the double-point defect is marked with a dotted line.}
\end{center}
\end{figure}

Figure \ref{fig:profiles_degererated} shows the values of $|p|$
obtained by MST versus the measured experimentally. The
experimental results are in very good agreement with the obtained
with MST. Note the different values of $|p|$ in the midpoint. As
MST predicts, a maximum is observed for the symmetric  mode at
$\nu=895$Hz, Figure \ref{fig:profiles_degererated}b  and a minimum
for the antisymmetric mode at $\nu=940$Hz, Figure
\ref{fig:profiles_degererated}a. The good agreement between
theoretical (MST) and experimental results is remarkable. These
measurements constitute the first experimental evidence of the
symmetric and antisymmetric vibrational modes inside the SC with a
double point defect.

\subsection{Decay of the localized modes}

Due to the mode of a single point defect is a little above of
the midgap, one can observe in Figure \ref{fig:EPWE_figure}b that
the localized modes in a double point defect present different
imaginary part of $k$: The values of the imaginary part of $k$ for
the antisymmetric mode are lower than the corresponding for the
symmetric mode, i.e., the rate of decay outside the cavity of the
symmetric mode must be bigger than that of the antisymmetric case.
In Figure \ref{fig:profiles_degererated} one can observe the decay
of the localized modes outside the double-point cavity. The border
of the double point defect are marked by dotted lines.

In order to analyze the decay of the modes outside the cavity, we
have studied the behaviour of the maximum analytical values of
$|p|$ outside the cavity (see open squares in Figure
\ref{fig:profiles_degererated}), calculated by MST in a SC of
rigid cylinders with size $11a\times 5a$. Although the decay of
the modes outside the cavity is multiexponential  \cite{Engelen09},
we can fit this values to an exponential-like $ae^{bx}$ for
analyzing the differences in the rate of decay due to the
differences in the imaginary part of the $k$ for each localized
modes.

\begin{figure}
\begin{center}
\includegraphics[width=90mm,height=50mm,angle=0]{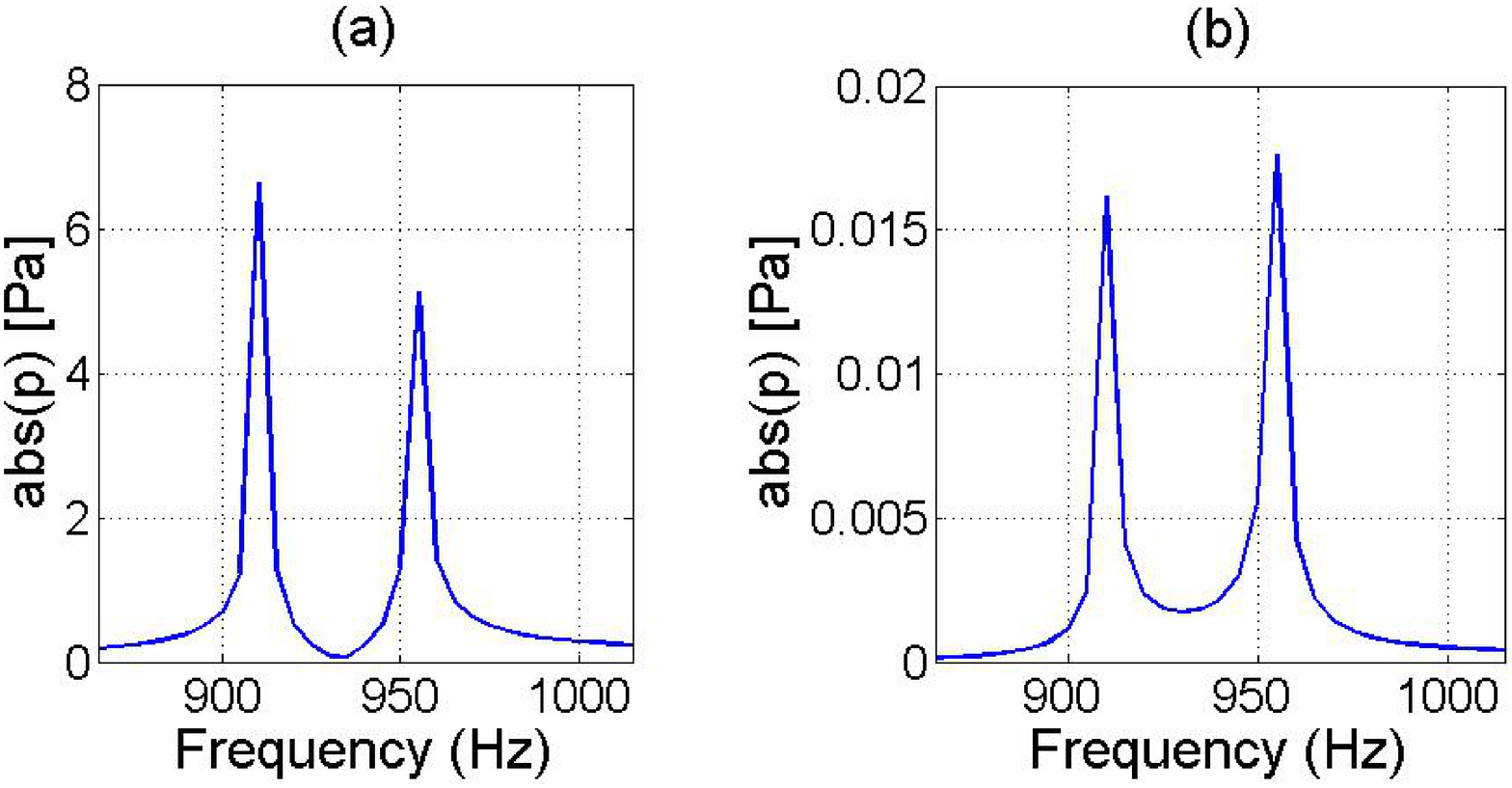}%
\caption{\label{fig:spectra} (Color Online) Spectra for a SC made
of PVC $9a\times 5a$ cylinders with a lattice constant $a=0.22$m
in square array with a double-point defect. (a) Spectrum measured
inside one of the point-defect in the double point defect. (b)
Spectrum measured outside the crystal at a distance $10a$ from the
start of the SC.}
\end{center}
\end{figure}

Both fitted exponential decays are represented in the Figure
\ref{fig:profiles_degererated} (dashed lines). The decay rate for
the antisymmetric mode is $b=-6.229\pm 0.237$ m$^{-1}$, while the
decay rate for the symmetric mode is $b=-6.633\pm 0.178$ m$^{-1}$.
Thus, as it was discussed in the results obtained in previous
section by EPEWE with supercell approximation, because of the
symmetric distribution of the frequencies of the localized modes
in double point defect respect to the localized mode in a single
cavity, the decay rate of the antisymmetric mode in a double-point
defect must be lower than that of the symmetric mode. On the other
hand, we can observe that the values of the decay rate of the
symmetric and of the antisymmetric modes are similar and the
difference between them is small. The splitting of the frequencies
of the localized modes in a double point defect around the
frequency of the single cavity , implies that the rate of decays
in double point defect have to be different, but also
 one of them should be smaller than the other because its distance to the center of the gap is bigger.

In finite crystals, where the localized modes can travel outside
the periodic structure because of their evanescent behaviour, the
previous results indicate that the symmetric mode will be killed
more easily than the antisymmetric mode. Thus, the design of
filters based on SC with point defect should take into account
this kind of results. In Figure \ref{fig:spectra}a we represent
the spectrum inside a point defect in a SC with a double-point
defect. In this case, we can observe that the value of pressure of
the peak of the symmetric mode (lower frequency) is higher than
the peak of the antisymmetric mode. In Figure \ref{fig:spectra}b
we represent the spectra for the same SC with a double-point
defect but measured outside the SC at a distance $10a$ from the
beginning of the SC. We can observe that the values of pressure
for the symmetric mode are lower than the values of the
antisymmetric mode, thus the symmetric mode has been more damped
by the crystal out of the double-point defect than the
antisymmetric mode. Similar results can be observed in Figure
\ref{fig:maps_degenerated}, where the acoustic field behind the
crystal for the antisymmetric mode is a bit bigger than the
corresponding for the symmetric mode. These results are in
complete agreement with the differences in the imaginary part of
$k$. Moreover, the difference in the value of imaginary of complex
wave vector is a direct evidence of the existence of different
vibrational modes in multi-point defects and revels the existence
of a coupling between them.

\section{Conclusions}
\label{sec:Conclusions}

Usual calculations with Plane Wave Expansion, $\omega(k)$ methods,
do not provide any information about the evanescent behaviour of
the localized modes in point defects inside periodic structures.
In this work we present the explicit matrix formulation of the
Extended Plane Wave Expansion (EPWE) with the supercell
approximation for a supercell with $N_p$ point defects for SC.
This technique allows us to study the evanescent behaviour of the
modes inside SC with multipoint defects. Localized modes in SC are
mainly characterized by three properties: splitting of
frequencies, the symmetry of the vibrational patterns and
evanescent behaviour inside the crystal. EPWE in addition to PWE
both with supercell approximation have been used to analyze the
whole properties of the localized modes in a SC with a double
point defect. First we analyze the splitting produced by the
generation of a double-point defect, showing the effects in both
the real and imaginary band structures. From the imaginary complex
band structure we obtain that the localized modes present
different values for the imaginary part of $k$, that means each
mode has a different decay rate inside the crystal. This property
has been observed experimentally by fitting the exponential decay
for each localized mode inside the crystal. The symmetry of the
vibrational patterns in double-point defect have also been
analyzed in the paper by means of MST calculation and experimental
data. Novel experimental evidences show the symmetric and
antisymmetric vibrational patterns in SC with double-point
defects. These data are in very good agreement with analytical
data. Finally, using the different decay rate of both vibrational
modes, we confirm our conclusions giving a new methodology to
determine different vibrational modes in periodic media. This work
shows the basis for the correct understanding of the design of
narrow filters and wave guides based on periodic structures with
multi-point defects.

\ack This work was supported by MEC (Spanish Government) and FEDER
funds, under grands MAT2009-09438 and MTM2009-14483-C02-02.

\section*{References}
\providecommand{\newblock}{}


\begin{thebibliography}{10}
\expandafter\ifx\csname url\endcsname\relax
  \def\url#1{{\tt #1}}\fi
\expandafter\ifx\csname urlprefix\endcsname\relax\def\urlprefix{URL }\fi
\providecommand{\eprint}[2][]{\url{#2}}

\bibitem{Sigalas93}
Sigalas M and Economou E 1993 {\em Solid State Commun.\/} {\bf 86} 141

\bibitem{Kushwaha93}
Kushwaha M, Halevi P, Dobrzynski L and Djafari-Rouhani B 1993 {\em Phys. Rev.
  Lett.\/} {\bf 71} 2022--2025

\bibitem{Yablonovitch}
Yablonovitch E 1987 {\em Phys. Rev. Lett.\/} {\bf 58} 2059

\bibitem{John87}
John S 1987 {\em Phys. Rev. Lett.\/} {\bf 58} 2486

\bibitem{Martinez05}
Mart\'inez-Sala R, Sancho J, S\'anchez-P\'erez J~V, G\'omez V, Llinares J and Meseguer
  F 1995 {\em nature\/} {\bf 378} 241

\bibitem{Khelif04}
Khelif A, Wilm M, Laude V, Ballandras S and Djafari-Rouhani B 2004 {\em Phys.
  Rev. E\/} {\bf 69} 067601

\bibitem{Joannopoulus08}
Joannopoulus J~D, Johnson S~G, Winn J~N and Meade R~D 2008 {\em Photonic
  Crystals. Molding the Flow of Light\/} (Princeton University press)

\bibitem{Sigalas05}
Sigalas M, Kushwaha M~S, Economou E~N, Kafesaki M, Psarobas I~E and Steurer W
  2005 {\em Z. Kristallogr.\/} {\bf 220} 765--809

\bibitem{Sigalas97}
Sigalas M 1997 {\em J. Acoust. Soc. Am.\/} {\bf 101} 1256

\bibitem{Sigalas98}
Sigalas M 1998 {\em J. Appl. Phys.\/} {\bf 84} 3026

\bibitem{Sanchez02}
S\'anchez-P\'erez J~V, Rubio C, Mart\'inez-Sala R, S\'anchez-Grandia R and
  G\'omez V 2002 {\em Appl. Phys. Lett.\/} {\bf 81} 5240

\bibitem{Vasseur08}
Vasseur J~O, Deymier P~A, Djafari-Rouhani B, Pennec Y and Hladky-Hennion A~C
  2008 {\em Phys. Rev.B\/} {\bf 77} 085415

\bibitem{Kushwaha94}
Kushwaha M, Halevi P, MartÌnez G, Dobrzynski L and Djafari-Rouhani B 1994 {\em
  Phys. Rev. B\/} {\bf 49} 2313--2322

\bibitem{Halevi95}
HHern\'andez-Cocoletzi, AKrokhin and PHalevi 1995 {\em Phys. Rev B\/} {\bf 51}
  17181--17183

\bibitem{Li05}
Li X and Liu Z 2005 {\em Solid State Communications\/} {\bf 133} 397ñ402

\bibitem{Wu01}
Wu F, Hou Z, Liu Z and Liu Y 2001 {\em Phys. Lett. A\/} {\bf 292} 198

\bibitem{Zhao09}
Zhao Y and LBYuan 2009 {\em J. Phys. D: Appl. Phys.\/} {\bf 42} 015403

\bibitem{Engelen09}
Engelen R, Mori D, Baba T and Kuipers L 2009 {\em Phys. Rev. Lett.\/} {\bf 102}
  023902

\bibitem{Romero10}
Romero-Garc\'ia V, S\'anchez-P\'erez J V, Casti\~neira Ib\'a\~nez S and Garcia-Raffi L M
2010 {\em Appl. Phys. Lett.\/} {\bf 96} 124102

\bibitem{Hsue05}
Hsue Y, Freeman A and Gu B 2005 {\em Phys. Rev B\/} {\bf 72} 195118

\bibitem{Laude09}
Laude V, Achaoui Y, Benchabane S and Khelif A 2009 {\em Phys. Rev. B\/} {\bf
  80} 092301

\bibitem{Romero10_2}
Romero-Garc\'ia V, S\'anchez-P\'erez J V and Garcia-Raffi L M 2010 {\em
  arXiv:1001.3758\/}

\bibitem{Wang09}
Wang Y, Li F, Kishimoto K, Wang Y and Huang W 2009 {\em Eur. Phys. J. B.\/}
  {\bf 67} 501--505

\bibitem{Sainidou05}
Sainidou R, Stefanou N, Psarobas I and Modinos A 2005 {\em Computer Physics
  Communications\/} {\bf 166} 197--240

\bibitem{Psarobas00}
Psarobas I~E, Stefanou N and Modinos A 2000 {\em Phys. Rev. B\/} {\bf 62} 278

\bibitem{Sainidou06}
Sainidou R and Stefanou N 2006 {\em Phys. Rev. B\/} {\bf 73} 184301

\bibitem{Wu03}
Wu F, Zhong H, Zhong S, Liu Z and Liu Y 2003 {\em Eur. Phys. J. B\/} {\bf 34}
  265--268

\bibitem{Zhong05}
Zhong H, Wu F, Zhang X and Liu Y 2005 {\em Physics Letters A\/} {\bf 339}
  478--487

\bibitem{Wu09PB}
Wu L, Chen L and Liu C 2009 {\em Physica B\/} {\bf 404} 1766

\bibitem{Wu09PLA}
Wu L~Y, Chen L~W and Liu C~M 2009 {\em Phys. Lett. A\/} {\bf 373} 1189--1195

\bibitem{Zhao09_2}
Zhao Y~C, Wu Y~B and Yuan L~B 2009 {\em Phys. Scr.\/} {\bf 80} 065401

\bibitem{Khelif03}
Khelif A, Choujaa A, Djafari-Rouhani B, Wilm M, Ballandras S and Laude V 2003
  {\em Phys. Rev. B\/} {\bf 68} 214301

\bibitem{Russell03}
Russell P~S~J, Marin E, DÌez A, Guenneau S and Movchan A~B 2003 {\em Opt.
  Expr.\/} {\bf 11} 2555

\bibitem{Linton01}
Linton C and McIver P 2001 {\em Handbook of Mathematical Techniques for
  Wave/Structure Interactions\/} (Chapman and Hall/CRC (Florida))

\bibitem{Chen01}
Chen Y~Y and Ye Z 2001 {\em Phys. Rev. E\/} {\bf 64} 036616 

\end{thebibliography}
\end{document}